\documentclass[12pt]{iopart}
\usepackage{graphicx}
\usepackage{gensymb}
\usepackage{cite}
\usepackage{latexsym}
\usepackage{amssymb}
\begin{document}

\title[Magnon dressing by orbital excitations in spin-orbital systems]
{Magnon dressing by orbital excitations in ferromagnetic planes
of K$_2$CuF$_4$ and LaMnO$_3$}

\author{Mateusz Snamina$^1$ and Andrzej M. Ole\'s$^{2,3}$}

\address{$^1$ Kazimierz Gumi\'nski Department of Theoretical Chemistry, Faculty of Chemistry,
              Jagiellonian University, Gronostajowa 2, PL-30387 Krak\'ow, Poland }
\address{$^2$ Max Planck Institute for Solid State Research,
              Heisenbergstrasse 1, D-70569 Stuttgart, Germany }
\address{$^3$ Marian Smoluchowski Institute of Physics, Jagiellonian University,
              Prof. S. \L{}ojasiewicza 11, PL-30348 Krak\'ow, Poland }

\ead{a.m.oles@fkf.mpg.de}

\vspace{10pt}
\begin{indented}
\item[]received 5 September 2018, revised 1 November 2018
\end{indented}

\begin{abstract}
We show that even when spins and orbitals disentangle in the ground
state, spin excitations are renormalized by the local tuning of $e_g$
orbitals in ferromagnetic planes of K$_2$CuF$_4$ and LaMnO$_3$.
As a result, dressed spin excitations (magnons) obtained within the
electronic model propagate as quasiparticles and their energy
renormalization depends on momentum ${\vec k}$. Therefore magnons in
spin-orbital systems go beyond the paradigm of the effective Heisenberg
model with nearest neighbor spin exchange derived from the ground state
--- spin-orbital entanglement in excited states predicts large magnon
softening at the Brillouin zone boundary, and in case of LaMnO$_3$ the
magnon energy at the $M=(\pi,\pi)$ point may be reduced by $\sim 45$\%.
In contrast, simultaneously the stiffness constant near the Goldstone
mode is almost unaffected. We elucidate physics behind magnon 
renormalization in spin-orbital systems and explain why long wavelength 
magnons are unrenormalized while simultaneously energies of short 
wavelength magnons are reduced by orbital fluctuations. In fact, the
${\vec k}$-dependence of the magnon energy is modified mainly by
dispersion which originates from spin exchange between second neighbors
along the cubic axes $a$ and $b$.
\end{abstract}


\submitto{\NJP}

\maketitle

\section{Introduction}

In $3d$ transition metal compounds strong intraorbital Coulomb
interaction $U$ leads to a Mott (or charge-transfer) insulator. Charge
excitations between two neighboring $3d$ ions with $m$ electrons per
site, $d_i^md_j^m\rightleftharpoons d_i^{m+1}d_j^{m-1}$, that
occur due to finite kinetic energy $\propto t$, generate superexchange
interactions $\propto J=4t^2/U$ \cite{Kho14}. In their pioneering work
Kugel and Khomskii \cite{Kug82} have shown that when degenerate orbitals
are partly filled, spin-orbital superexchange couples spin and orbital
degrees of freedom. It leads to phases with spin-orbital superexchange
in two-dimensional (2D) \cite{Ver04,Rei05,Cha08,Nor08,Cor11,Brz12}
or in three-dimensional (3D)
\cite{Oka02,Fei99,Miz99,Kim04,Ole05,Endoh,Zen05,Sol08,Jan09,Brz13,Fuj13,Wu13}
systems. When both spin and orbital degrees of freedom are active
joint spin-orbital quantum fluctuations arise and may even destabilize
long-range order \cite{Fei97}. These fluctuations are the strongest
for $t_{2g}$ orbital degrees of freedom \cite{Kha05}, where the spin
exchange derived from spin-orbital superexchange is strongly entangled
and has a dynamical character \cite{Ole06,Ole12}. In model systems
spin-orbital entanglement may be used to identify quantum phase
transitions \cite{You15}.

Orbital degeneracy opens the route towards complex types of
spin-orbital order with coexisting antiferromagnetic (AF) and
ferromagnetic (FM) exchange bonds. Frequently such systems are
analyzed using the classical Goodenough-Kanamori rules \cite{Goode}
which emphasize the complementarity of spin and orbital order, i.e.,
alternating orbital (AO) order supports FM spin exchange and
ferro-orbital (FO) order supports AF exchange. They follow from the
assumption that spin and orbital excitations are independent of each
other and spin exchange interactions may be derived from the
spin-orbital superexchange by averaging over the orbital state. Indeed,
when joint spin-orbital fluctuations are quenched, e.g. by lattice
distortions, these rules apply and the disentangled superexchange
helps to understand experimental observations \cite{Ole05}.
A good example of this approach is the parent compound of colossal
magnetoresistance manganites LaMnO$_3$ \cite{Dag01}, with small
spin-orbital entanglement \cite{Sna16}. Therefore, spin waves measured
in inelastic neutron scattering \cite{Mou96,Mur98,Kar08} and optical
spectral weights \cite{Kov10} could be successfully interpreted using
the effective anisotropic Heisenberg model.

In doped manganites double exchange provides a large FM exchange
interaction \cite{Dag01}. It is responsible for the onset of FM order
and modifies occupied $e_g$ orbitals involved in the hopping process as
demonstrated in the one-dimensional (1D) spin-orbital model
\cite{Dag04}. Hole-orbital and orbital-lattice fluctuations were
identified as the main origin of the observed unusual softening of the
magnon spectrum at the zone boundary \cite{Kha00,Sin10,Sin13}. It has
been shown that orbitons depend thereby on magnons in Mott insulators
with orbital degrees of freedom \cite{Tan04,Sch12,Woh11,Kim12},
and both contribute to spectral properties \cite{Ish05,Mona}.

In 1D cuprates \cite{hiha} (2D iridates \cite{haha})
orbitons (excitons) are dressed by magnons, while the opposite effect
of orbital excitations on magnons was considered only in the context
of the strong zone boundary magnon softening observed experimentally
in manganites close to half doping \cite{Sin10}.
In this paper we demonstrate that spin excitations in a FM plane with
AO order, as in K$_2$CuF$_4$ \cite{Mor04} or LaMnO$_3$ \cite{Tok06},
are indeed renormalized by the changes of occupied $e_g$ orbitals,
leading to magnons dressed by orbital fluctuations and propagating
together as a quasiparticle in a Mott insulator. This phenomenon is
similar to the local changes of AF spin order by an added hole in
superconducting cuprates \cite{Lau16}.

The remaining of the paper is organized as follows. In section 2 we
introduce a general form of spin-orbital Hamiltonian with $e_g$
degrees of freedom and present the magnon excitations starting from
orbital order in the ground state. Next we release the constraint of
frozen orbitals and present the variational way of finding magnon
excitations for optimized orbitals in section 3. A simplified version
of this approach and a numerical \textit{Ansatz} which serves to
verify the predictions of the variational approach are presented in
section 4. The results for magnons in K$_2$CuF$_4$ are given in
section 5. We consider a spin excitation in the FM planes of
LaMnO$_3$ and analyze the optimal orbital angles near the excitation
in section 6. There we also show that the effective spin model will
include nearest neighbor $J_1$, next-nearest neighbor $J_2$, and third
next neighbor $J_3$ spin exchange, although the spin-orbital
superexchange couples only nearest neighbors.
Analytic estimation of the renormalized interaction $J_1$ which
determines the magnon bandwidth is presented in the Appendix.
The paper is concluded with a short summary in section~7.

\section{Spin-orbital model and magnons for frozen orbitals}

We begin with the $e_g$ orbital basis
(labeled in analogy to $|{\uparrow}\rangle$ and $|{\downarrow}\rangle$
spin $S=\frac12$ states):
\begin{equation}
\label{real}
|\zeta_c\rangle\equiv\frac{1}{\sqrt{6}}\left(3z^2-r^2\right),
\hspace{0.7cm}
|\xi_c\rangle  \equiv\frac{1}{\sqrt{2}}\left(x^2-y^2\right),
\end{equation}
i.e., a directional orbital $|\zeta_c\rangle$ along the $c$ axis,
and an orthogonal to it planar orbital $|\xi_c\rangle$.
The energetic splitting of $e_g$ states,
\begin{equation}
\label{Hz}
\hat H_z=\frac12
         \sum_i\left(|i\zeta_c\rangle\langle i\zeta_c|
                    -|i\xi_c\rangle\langle i\xi_c|\right)
        =\sum_i\hat\tau_i^{(c)},
\end{equation}
selects the favored orbital at site $i$ by the tetragonal crystal
field $\propto E_z$.
We consider a generic 2D $e_g$ spin-orbital superexchange model on
a square lattice,
\begin{equation}
\label{som}
\hat{\cal H} = J\left(
c_1\hat H_1+c_2\hat H_2+c_3\hat H_3\right)+E_z\hat H_z,
\end{equation}
which explains FM order of spins $S$ in the $ab$ planes of K$_2$CuF$_4$
($S=\frac12$) \cite{Ole00} or LaMnO$_3$ ($S=2$) \cite{Fei99}, with
three $\{{\hat H}_n\}$ terms explained below. The positive coefficients
\mbox{$\{c_1,c_2,c_3\}$} depend on the multiplet structure of excited
 $3d^8$ Cu$^{3+}$ states \cite{Ole00}
($3d^5$ Mn$^{2+}$ states \cite{Fei99})
via Hund's exchange $J_H/U$ \cite{Ole05}. In the ground state an $e_g$
hole at a Cu$^{2+}$ ion in K$_2$CuF$_4$ (an $e_g$ electron at a
Mn$^{3+}$ ion in LaMnO$_3$) occupies a linear combination of two
orbital states (\ref{real}) at site $i$ \cite{Kho14},
\begin{equation}
 |i\vartheta\rangle\equiv
     \cos(\vartheta/2) \left|i\zeta_c\right\rangle
   + \sin(\vartheta/2) \left|i\xi_c  \right\rangle.
\label{mixing}
\end{equation}
When tetragonal distortion is ignored ($E_z=0$), the occupied orbital
states on two sublattices $A$ and $B$ are symmetric/antisymmtric
combinations of $e_g$ orbital basis $\{|\zeta_c\rangle,|\xi_c\rangle\}$
states with $\vartheta=\pm\pi/2$, otherwise for positive (negative)
values of $E_z$, enhanced amplitude of $|x^2-y^2\rangle$
$\left(|3z^2-r^2\rangle\right)$ orbital states is favored.

The superexchange part $\propto J$ in Eq. (\ref{som}) involves spin
$\{\hat{\vec S}_i\}$ and orbital pseudospin
$\left\{\hat\tau_i^{(\gamma)}\right\}$ ($\gamma=a,b,c$) operators ---
it consists of three terms which follow from high-spin
(${\hat H}_1$), low-spin interorbital (${\hat H}_2$), and low-spin
intraorbital (${\hat H}_3$) charge excitations along nearest neighbor
$\langle ij\rangle$ bonds,
\begin{eqnarray}
\label{som1}
\hat H_{1}
 &=&
 -\sum_{\gamma} \sum_{\langle ij \rangle\parallel\gamma}\!
 \left[\hat{\vec S}_i\!\cdot\!\hat{\vec S}_j+ S(S\!+\!1)\right]
 \otimes
\left(\frac14 - \hat\tau_i^{(\gamma)}\hat\tau_j^{(\gamma)}\right),
 \\
\label{som2}
\hat H_{2}
 &=&
 \sum_{\gamma}\sum_{\langle ij \rangle\parallel\gamma}\!
 \left(\hat{\vec S}_i\!\cdot\!\hat{\vec S}_j - S^2 \right)
 \otimes
\left(\frac14 - \hat\tau_i^{(\gamma)}\hat\tau_j^{(\gamma)}\right),
 \\
\label{som3}
\hat H_{3}
 &=&
 \sum_{\gamma}\!\sum_{\langle ij \rangle\parallel\gamma}\!
 \left(\hat{\vec S}_i\!\cdot\!\hat{\vec S}_j - S^2 \right)
 \otimes
\left(\frac12 - \hat\tau_i^{(\gamma)}\right)
  \left(\frac12 - \hat\tau_j^{(\gamma)}\right).
\end{eqnarray}
The orbital operators $\hat\tau_i^{(a)}$ and $\hat\tau_i^{(b)}$ for $ab$
planes follow from $\hat\tau_i^{(c)}$ along the $c$ axis by a cubic
transformation \cite{Ole00}, see Eq. (\ref{Hz}).

In K$_2$CuF$_4$ one finds FM order at $J_H/U\sim 0.2$ \cite{Oni08},
coexisting with AO order \cite{Ito76} of hole orbital states with
angles $\pm\theta_{opt}$ on the two sublattices, $A$ and $B$. Averaging
the orbital operators over this AO order in $ab$ planes gives spin
Hamiltonian with FM exchange (with $J_{\lozenge}>0$),
\mbox{
$H=-J_{\lozenge}\sum_{\langle ij\rangle}\hat{\vec S}_i\cdot\hat{\vec S}_j$
}, which served to interpret the experimental data \cite{Mou76,Hir83}.
To investigate magnons (spin waves) we create a spin excitation at site
$i=0$ by decreasing the value of $S_0^z=S$ to $S_0^z=(S\!-\!1)$. In the
simplest approach we disentangle \cite{Ole12} spin-orbital superexchange
both in the ground and in excited states and use the same frozen AO
order shown in Fig. \ref{fig:orbs}(a) to determine spin exchange
$J_{\lozenge}$.

A spin excitation (a magnon) itself is best described by the
transformation to Holstein-Primakoff (HP) bosons \cite{HoP}.
In the linear spin-wave theory magnon energy consists of two
contributions and we introduce:\\
($i$) Ising energy for a localized HP boson
$I^{(0)}\equiv 4J_{\lozenge} S$ and \\
($ii$) the propagating term
$P^{(0)}({\vec k})\equiv -4J_{\lozenge} S\gamma_{\vec k}$. \\
The latter originates from quantum fluctuations
$\propto -\frac12 J_{\lozenge}
({\hat S}_i^+{\hat S}_j^-+{\hat S}_i^-{\hat S}_j^+)$,
where
$\gamma_{\vec k}=\frac14\sum_{\vec\delta}e^{i{\vec k}\cdot{\vec\delta}}$
depends on the momentum ${\vec k}=(k_a,k_b)$ with
$k_\alpha\in[-\pi,\pi)$.
Here $\vec\delta$ stands for one of four nearest neighbors of the
central site $i=0$ shown in Fig. \ref{fig:orbs}(a). The above two
terms determine the magnon dispersion in a 2D ferromagnet,
\begin{equation}
\omega_{\vec k}^{(0)}=I^{(0)}+P^{(0)}({\vec k})
=4J_{\lozenge}S(1-\gamma_{\vec k}),
\label{omega}
\end{equation}
which serves as a reference below.
The breaking of SU(2) symmetry is reflected by a Goldstone mode
(at ${\vec k}=0$), and $\omega_{\vec k}=J_{\lozenge}S k^2$ for
${\vec k}\to 0$ --- we find that this result is insensitive to
spin-orbital coupling.

In general however orbitals are not frozen in a spin-orbital system
and will respond locally to a spin excitation. One might expect that
this reduces spin exchange, $J_{\lozenge}\rightarrow J_{\blacklozenge}$
and the magnon dispersion would soften. Indeed, we have found that the
magnon energy $\omega_{\vec k}^{(0)}$ is reduced but this effect is
rather subtle and the renormalization of exchange interaction
$J_{\blacklozenge}$ depends on momentum ${\vec k}$.

\section{Variational approximation}

\begin{figure}[t!]
\begin{center}
\includegraphics[width=15.4cm]{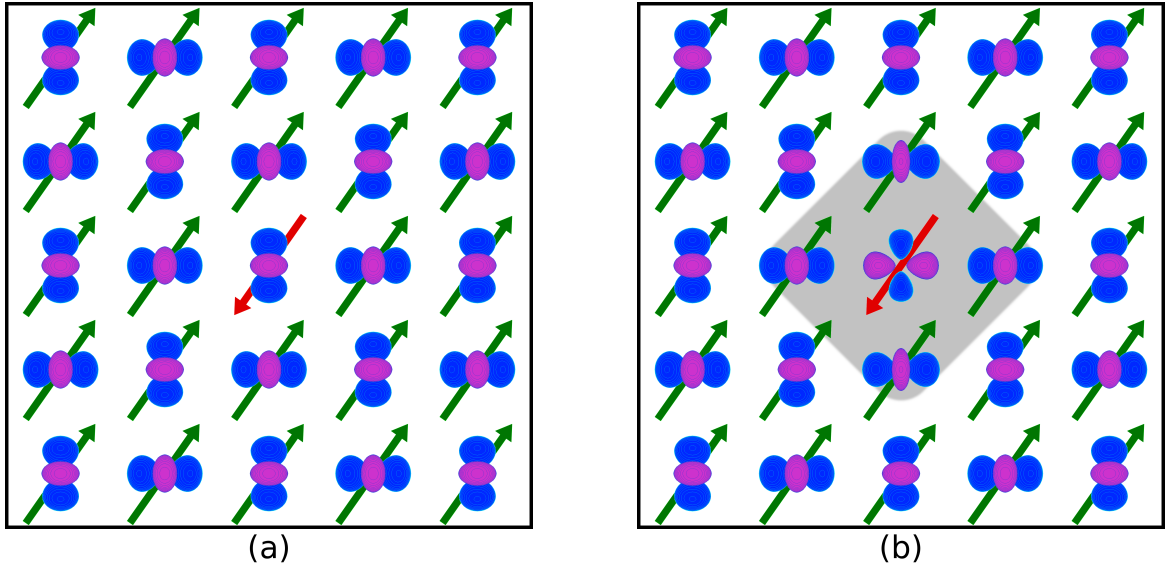}
\end{center}
\protect\caption{Artist's view of a spin excitation (inverted red
arrow) in the FM plane of K$_2$CuF$_4$ (green arrows) and AO order
(of hole orbitals) at $E_z=-0.8J$, with:
(a)~frozen orbitals;
(b)~optimized orbitals at the central spin-flip site itself and at four
its neighboring sites in the square lattice, forming a quasiparticle
(dressed magnon). The above value of $E_z$ leads to the expected AO
order in K$_2$CuF$_4$ \cite{Ito76}
(different colors indicate the orbital phases), with
$\theta_{opt}\simeq 71{\degree}$ in Eq. (\ref{mixing}).
When the VA is used, case (a) is still realized at $\vec{k}\simeq 0$,
while case (b) represents a dressed magnon with $\vec{k}\simeq M$
where orbital states in the shaded cluster are radically different
from those in the ground state, cf. frozen orbitals in (a).
}
\label{fig:orbs}
\end{figure}

To capture the response of orbital background to a spin excitation
we invoke the following Variational Approximation (VA): Significant
changes of occupied orbitals with respect to the reference AO order
are expected at the nearest neighbors of excited spin and at the site
of spin excitation itself, see Fig. \ref{fig:orbs}(b). The largest
change at sublattice $L=A,B$, $\lambda_L(\theta_{opt}+\theta_{1L})$
with $\lambda_A=1=-\lambda_B$, occurs at the site of spin excitation
itself. For the neighboring sites we use the lattice symmetry and
search for the same optimal orbitals given by angles
$-\lambda_L(\theta_{opt}+\theta_{2L})$ and
$-\lambda_L(\theta_{opt}+\theta_{3L})$ at equivalent neighbors along
each cubic axis, $a$ or $b$.

It is crucial that the VA is performed for each value of momentum
${\vec k}$ independently. We have evaluated the matrix elements of the
Hamiltonian $\hat{\cal H}$ (\ref{som}) for a single spin excitation in
the thermodynamic limit, and determined six variational parameters
$\{\theta_{iL}\}$ ($i=1,2,3$; $L=A,B$). In this way we obtained
the renormalized magnon dispersion which replaces Eq. (\ref{omega}),
\begin{equation}
\omega_{\vec k}(\{\theta_{iL}\})=
I(\{\theta_{iL}\};{\vec k})+P(\{\theta_{iL}\};{\vec k}).
\label{dE}
\end{equation}
By construction the angles $\{\theta_{iL}\}$ are real as in Eq.
(\ref{mixing}); we have verified that complex coefficients do not lead
to further significant energy lowering.

The AO order has a unit cell consisting of two atoms which defines the
reduced Brillouin zone (RBZ). The magnon dispersion consists then of
two branches in the RBZ, the lower one for $|k_a|+|k_b|\le\pi$, i.e.,
${\vec k}\!\!\in\,$RBZ, and the upper one for vectors
$({\vec k}+{\vec Q})\!\!\notin\,$RBZ, where ${\vec Q}=(\pi,\pi)$ is the
nesting vector. The two magnon branches give a gapless dispersion and
are determined in two steps to take the full advantage of variational
parameters. First we find the magnon energies from the lower magnon
band --- they depend on $I_A(\{\theta_{iA}\};{\vec k})$,
$I_B(\{\theta_{iB}\};{\vec k})$ and $P_{AB}(\{\theta_{iL}\};{\vec k})$.
The terms $I_L(\{\theta_{iL}\};{\vec k})$ stand for the Ising HP boson
parts, while $P_{AB}(\{\theta_{iL}\};{\vec k})$ (with $L=a,b$) is
obtained from the HP boson propagation along the bonds parallel to the
$a$ and $b$ axis, from sublattice $A$ to $B$ (or vice versa).

The eigenstates of the second magnon band are determined in the second
step --- magnon states which momenta do not belong to the RBZ.
As a magnon states with momentum ${\vec k}\!\notin$RBZ is orthogonal to
its partner magnon state with momentum $({\vec k}-{\vec Q})$, then, at
this stage, the variational principle
has to be applied together with rigorous orthogonality condition.

\section{Simplified variational approximation and numerical \textit{Ansatz}}

Assuming that orbital optimization for both sublattices is equivalent,
we use the constraint \mbox{$\theta_i\equiv\theta_{iA}=\theta_{iB}$}
($i=1,2,3$) which defines Simplified Variational Approximation (SVA).
Here we consider the full Brillouin zone and evaluate the energies of
a dressed HP boson $I(\{\theta_i\})$ and of its propagating part
$P(\{\theta_i\};{\vec k})$.
The SVA is equivalent to the VA when the magnon happens to be a
symmetric linear combination of the two waves propagating over the two
sublattices --- this concerns the $\Gamma-M$ direction; otherwise one
may expect that the amplitude of the spin wave is larger in one
sublattice and the magnon wave function differs qualitatively from
that obtained for a Heisenberg ferromagnet.
Below we show that the VA gives indeed better results than the SVA,
and the magnon dressing occurs differently on both sublattices.

Finally, we verified the predictions of the VA by exact diagonalization
employing a Numerical \textit{Ansatz} (NA) with six states per
sublattice: a spin defect with or without orbital excitation and four
spin-orbital states with spin excitation at the central site together
with an orbital excitation at one of nearest neighbors. The state with
excitations within a shaded cluster depicted in Fig. 1(b) may be thus
expressed in terms of these six states. Here the constraint for equal
orbital angles at two neighbors along the same axis is released.
The eigenstates and the spin excitation energy $\omega_{\vec k}$ are
found separately by exact diagonalization of a $12\times 12$ matrix
obtained for each momentum ${\vec k}$.

\section{Magnons for K$_2$CuF$_4$}

\begin{figure}[t!]
\begin{center}
\includegraphics[width=16cm]{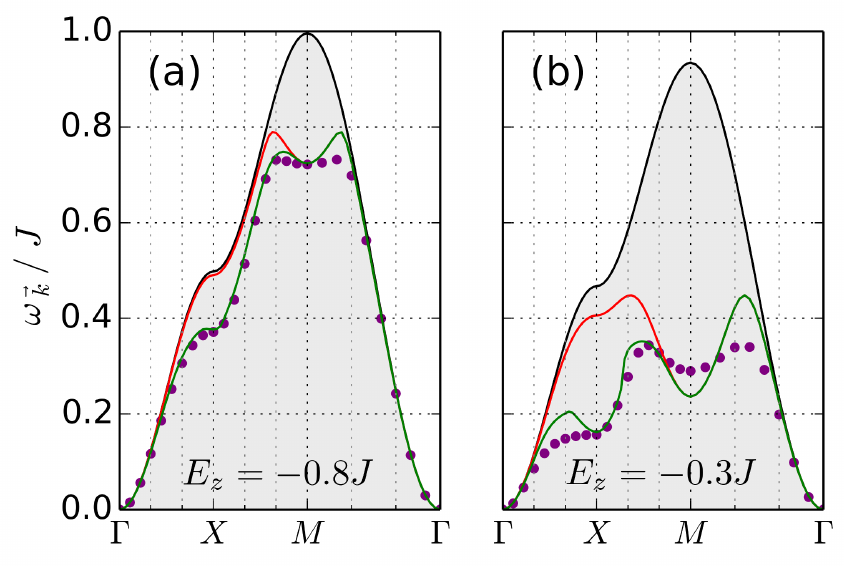}
\end{center}
\protect\caption{
The magnon energy $\omega_{\vec k}/J$ obtained for FM state of
K$_2$CuF$_4$ at $J_H/U=0.2$ and:
(a)~$E_z=-0.80J$ and
(b) $E_z=-0.30J$.
Results are presented for four approximations:
frozen orbitals (black line and grey background),
the VA (green line),
the SVA (red line), and
the 12-state NA (purple dots).
The~high symmetry points are:
$\Gamma\!=(0,0)$, $X\!=(\pi,0)$, $M\!=(\pi,\pi)$.
}
\label{fig:kcf}
\end{figure}

Taking as an example the K$_2$CuF$_4$ state at $E_z=-0.8J$ shown in
Fig. \ref{fig:orbs}(b), one finds that the orbital renormalization is
large --- at the central site with spin excitation it is largely
modified to \mbox{$\sim(x^2-y^2)$} and the orbitals at the four
neighboring sites are also changed. The latter orbitals found within
the VA are only weakly changed as these latter sites have three
neighbors belonging to the neighbors with undisturbed AO order
\cite{Ito76}, but the one at the spin excitation itself is radically
different. For this reason we introduce a cutoff and assume that the
orbitals at further neighbors of the excited spin are unchanged.
One expects then large dressing of the magnon, with the corresponding
reduction of the effective FM interaction to $J_{\blacklozenge}$,
particularly in the neighborhood of the $M$ point. This is confirmed
by the results shown in Fig. \ref{fig:kcf}(a) --- the magnon energy
$\omega_M$ is reduced by $\sim 27$\% from $\omega_M^{(0)}$.
Internal consistency of the theory is confirmed by this reduction
being nearly the same in all three methods treating spin-orbital
coupling: VA, SVA, and NA.

At the $X$ point we recognize the importance of independent
optimization of orbitals at the two sublattices ---
the energy $\omega_X$ is reduced by $\sim 25$\% from $\omega_X^{(0)}$
in the VA while it stays almost unrenormalized in the SVA, see Fig.
\ref{fig:kcf}(a). The NA agrees very well with the results of the VA
except for the points close to the $M$ point along the $M-\Gamma$ path.
While the VA may underestimate somewhat the magnon dressing effect,
altogether we find indeed a comparison of the VA with
the NA very encouraging. The renormalization of magnon energy
increases fast when the orbital splitting $|E_z|$ is reduced, and
one finds that the magnon energy reduction is large for $E_z=-0.3J$,
e.g. by $\sim 60$\% at the $M$ point, see Fig. \ref{fig:kcf}(b).
The agreement between the VA and the NA is somewhat worse here but
remains still in qualitative agreement.
Altogether, we suggest that the magnon softening may be very large for
spin-orbital systems with low spin $S=\frac12$ as in K$_2$CuF$_4$.

\section{Magnons for ferromagnetic planes of LaMnO$_3$}

\begin{figure}[t!]
\begin{center}
\includegraphics[width=16cm]{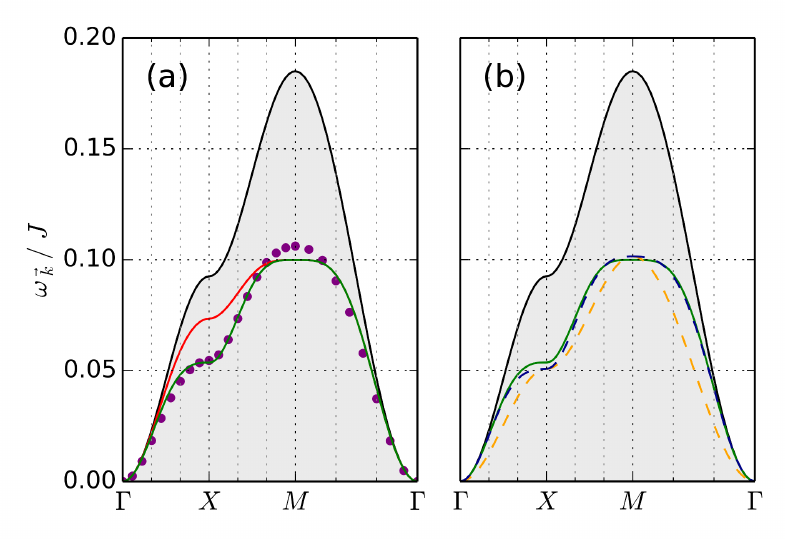}
\end{center}
\protect\caption{
The magnon energy $\omega_{\vec k}/J$ for the FM plane of LaMnO$_3$
at $E_z=+1.04J$ obtained:
(a) in various approximations, i.e., the
\textit{frozen orbital} approach, the VA, and the SVA (black, green,
and red line), and the NA (purple dots), and
(b) in the VA (green line) and fitted using the Heisenberg model with
nearest neighbor $J_1=6.34\times 10^{-3}J$ interaction only
(orange line), and with both the above nearest neighbor $J_1$ and
third nearest neighbor $J_3=1.35\times 10^{-3}J$ interactions
(dark blue line). Gray shading highlights the difference between the
\textit{frozen orbital} approach and the VA.
Parameter: $J_H/U=0.1725$ \cite{Fei99}.
}
\label{fig:lmo}
\end{figure}

For LaMnO$_3$ we consider electrons in $e_g$ orbitals at $E_z>0$
and use a representative value of the orbital spitting \cite{Sna18}
\mbox{$E_z=10c_1J\simeq 1.04J$}
which gives $\theta_{opt}\simeq 120{\degree}$. Spin and orbital
excitations depend on $\theta_{opt}$ for frozen orbitals \cite{Bal01}.
The magnon dispersion is modified within the VA or the SVA, see Fig.
\ref{fig:lmo}(a). In agreement with our initial intuition, the magnon
energies predicted for dynamical orbitals soften. The energy lowering
from $\omega_{\vec k}^{(0)}$ to $\omega_{\vec k}$ is substantial for
this value of $E_z$ --- up to about 45\% at the $M$ point. We emphasize
that the VA and the NA agree almost perfectly and this agreement
confirms \textit{a posteriori} our initial choice of real orbital
phases in the VA. One observes that the energy $\omega_{\vec k}$ is
somewhat lower than in the NA in the neighborhood of the $M$ point,
indicating that the orbitally doubly excited states become important
when at least two orbital deformations are large enough
(such states are not included in the NA).

We remark that the reported experimental spin exchange constants are
the final product of processing the experimental data concerning the
magnons energies. A link between them and the measured energies is
established by a parametrized form of the dispersion relation for some
conceived pure-spin models defined by a specific interactions pattern.
In case of LaMnO$_3$, the simplest Heisenberg model with nearest
neighbor interaction $J_1$ was successfully used to interpret the
experimental data in the past \cite{Mou96} ---
it predicts the magnon dispersion given by Eq. \eref{omega}.

We decided to follow the same strategy and studied our
\textit{calculated} magnon dispersion $\omega_{\vec k}$ in Fig.
\ref{fig:lmo}(a). We tested whether or not one may fit the calculated
magnon energies with the effective Heisenberg model and how many
exchange interactions are needed using the dispersion $\omega_{\vec k}$
discretized over a mesh of $(k_a,k_b)$ values. It turned out that to
reproduce the magnon bandwidth $8J_1S$ the fit requires nearest
neighbor exchange interaction $J_1=6.34\times 10^{-3}J$,
see Fig. \ref{fig:lmo}(b). Although the reduced value of the magnon
bandwidth is then reproduced, the ${\vec k}$-dependence of
$\omega_{\vec k}$ near the $M$ point is not. It is clear that the
Heisenberg model with nearest neighbor exchange is insufficient as the
obtained dispersion $\omega_{\vec k}$ deviates then from the one
derived from the VA, particularly near the $M$ point. The fit may be
refined by taking into account the next-nearest and third neighbor
interactions, $J_2$ and $J_3$, in the effective spin model. One finds
that $J_2=0.25\times{10}^{-3}J$ is rather small but
$J_3\simeq 1.35\times{10}^{-3}J$ is significant and plays an important
role. Both fits are shown in Fig. \ref{fig:lmo}(b).

The fitted value of the nearest neighbor exchange spin constant
$J_1=6.34\times 10^{-3}J$ is much smaller than the value
$J_\lozenge=11.56\times 10^{-3}J$ obtained in the
\textit{frozen orbital approach}, actually by $5.22\times 10^{-3}J$,
i.e., by $\sim 0.45J_\lozenge$. This reduction of $J_1$ may be
rationalized and was also calculated analytically using our SVA,
see the Appendix. Expanding the obtained dispersion $\omega_{\vec k}$
in the range of small ${\vec k}\to 0$, we derived that
\begin{equation}
J_\lozenge\simeq J_1+4J_3.
\label{gold}
\end{equation}
This explains why: \hfill\break
($i$) the overall magnon bandwidth of $8J_1S$ is here strongly reduced
from $\omega_M^{(0)}\simeq 0.185J$ to $\omega_M\simeq 0.101J$, but
simultaneously \hfill\break
($ii$) the stiffness constant determined by $J_1+4J_3$ (by $J_\lozenge$
for \textit{frozen orbitals}) remains unrenormalized \cite{Fer98},
see Eq. (\ref{gold}). \hfill\break
We emphasize that in this way an outstanding question in the theory how
these two effects may occur simultaneously \cite{Kha00} is explained.

\begin{figure}[t!]
\begin{center}
\includegraphics[width=16cm]{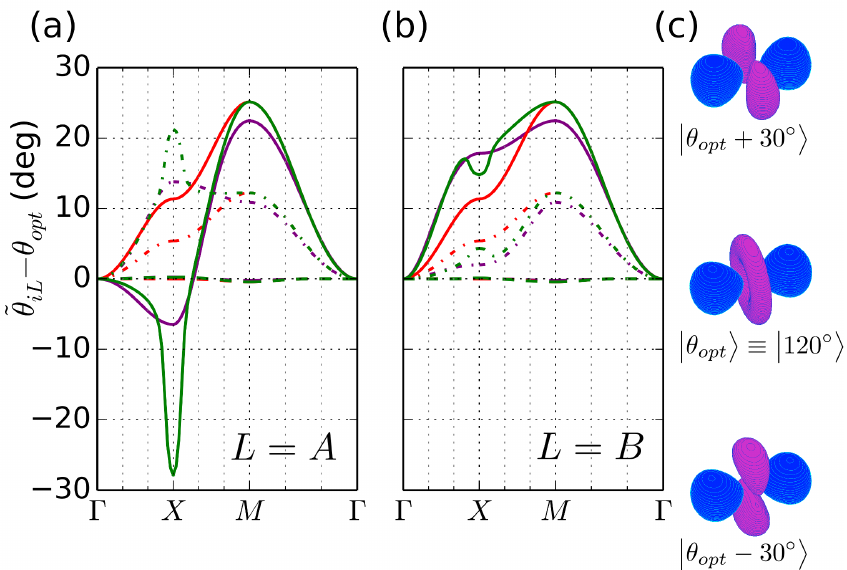}
\end{center}
\protect\caption{
The orbital changes at nearest (solid), next-nearest (dashed),
and third nearest neighbor (dot-dashed lines) of spin excitation in
LaMnO$_3$ found in the VA (green), the SVA (red), and the NA
(purple lines) for HP bosons at sublattice:
(a) $A$ and (b) $B$.
In (c) orbital states (\ref{mixing}) are shown (different colors stand
for different orbital phases) for:
$\theta_{opt}=120{\degree}$ (middle),
$\theta_{iL}=\theta_{opt}-30{\degree}$ (bottom), and
$\theta_{iL}=\theta_{opt}+30{\degree}$ (top).
Parameters as in Fig. \ref{fig:lmo}.
}
\label{fig:sub}
\end{figure}

Altogether, the lowering of magnon energy is quite similar for all the
methods although some discrepancies occur.
We observe that the SVA is here again insufficient when the magnon
momentum has large imbalance between its components $k_a$ and $k_b$.
Indeed, for $\vec k$ being close to the $X$ point, the SVA is able to
give only half of the magnon softening seen in the VA or in the
NA, see Fig. \ref{fig:lmo}. Good agreement between the VA and the NA
found here and in K$_2$CuF$_4$ at $E_z=-0.8J$ justifies
\textit{a posteriori} the idea of independent
determination of orbital angles for the sublattices $A$ and $B$.

The optimal orbital angles $\tilde\theta_{iL}$ for a magnon dressed by
orbital excitations are changed in LaMnO$_3$ by less than
$\pm 30{\degree}$ and remain quite similar to the ground state orbitals
with $\theta_{opt}=120{\degree}$, see Fig. \ref{fig:sub}. In general
orbital angles increase for the dressed HP bosons. This may be
explained because the optimal values of orbital angles
$\tilde\theta_{iL}$ follow from the interplay between superexchange
interaction and tetragonal crystal field~$E_z$. The first one favors
$\theta=90{\degree}$, while the second one favors $\theta=180{\degree}$.
When a HP boson is created, the spin exchange effectively decreases
while the value of $E_z$ is not affected.

\section{Conclusions}

Summarizing, we have shown that spin-orbital superexchange tunes the
orbital angles near local spin excitations and is responsible for novel
dressed magnon quasiparticles. The magnon-orbiton coupling is local and
reduces nearest neighbor spin exchange $J_1$ responsible for magnon
dispersion at the $M$ point, while orbital fluctuations couple
predominantly to spin excitations at neighboring sites and this
generates third nearest neighbor $J_3$ exchange couplings. Thus
spin-orbital entanglement has here similar consequences to the exotic
phase suggested as a possible ground state of the 2D Kugel-Khomskii
model \cite{Brz12}. There the effective spin model derived near a
quantum phase transition includes next nearest ($J_2$) and third 
nearest ($J_3$) neighbor exchange interactions and the latter are of 
crucial importance to stabilize spin orientation. Here spin-orbital 
entanglement generates also $J_3$ interactions which couple spins 
distant by two lattice constants along the cubic axes $a$ and $b$, 
and thus the magnon dispersion is different from that given by the 
Heisenberg model with nearest neighbor exchange constants derived 
from frozen orbitals when spin-orbital interactions are disentangled 
in the ground state. We suggest that such effects would be weaker but 
still measurable in 3D ordered phases with ferromagnetic planes, as 
for instance in LaMnO$_3$, and it is very challenging to detect them.

In the electronic model considered here spin-orbital degrees of 
freedom are entangled and thus respond jointly, giving renormalized 
spin excitations. However, strong coupling between orbitals and 
lattice distortions caused by the Jahn-Teller effect will reduce the
magnon-orbiton entanglement and thus the renormalization of magnon
dispersion reported here will decrease. We suggest that only future
experiments could establish importance of spin-orbital
entanglement in excited states.

We suggest that similar analysis of the magnon dispersion could be
performed using the variational approach for doped ferromagnetic
manganites with statistically averaged interactions between initial
$S=2$ spins at Mn$^{3+}$ ions and $S=3/2$ spins at Mn$^{4+}$ ions
\cite{Ole02}. We expect that it would
reproduce the reduction the magnon energies at the Brillouin zone
boundary obtained in the diagrammatic approach \cite{Sin10}.
Finally, we remark that the present variational method could also be
used to investigate magnon dispersion in the charge, orbital, and spin
ordered phase in La$_{1/2}$Sr$_{3/2}$MnO$_4$ \cite{Senff}.

\section*{Acknowledgments}

We thank Wojciech Brzezicki, Krzysztof Ro\'sciszewski, Krzysztof Wohlfeld,
and particularly Reinhard Kremer and Giniyat Khaliullin for insightful
discussions. We kindly acknowledge support by Narodowe Centrum Nauki
(NCN, Poland) under Project No.~2016/23/B/ST3/00839.
A M Ole\'s is grateful for the Alexander von Humboldt-Stiftung
Fellowship (Humboldt-Forschungspreis).

\section*{Appendix:
          Analytic estimation of the nearest neighbor exchange $J_1$}

Creation of magnons characterized by $\vec k=0$ (being Goldstone modes)
does not entail any changes in the orbital background for a spin-orbital
system. When magnons characterized by finite $\vec k\simeq 0$ are created,
the coupled orbitals may be slightly modified. As a result, $I$ and
$P({\vec k})$ terms deviate from $I^{(0)}$ and $P^{(0)}({\vec k})$.
To highlight the minute changes due to spin-orbital entanglement, we
introduce a vector $\mathbf{x}$ consisting of differences in variational
parameters with respect to their values in the ground state, and expand
$I$ and $P$ in terms of $\mathbf{x}$ treated as a small parameter:
\begin{eqnarray}
 I(\mathbf{x})&\simeq&
 I^{(0)}
 +
 \mathbf{u}^T\mathbf{x}
 +
 \frac{1}{2} \mathbf{x}^T \mathbf{U} \mathbf{x},
 \\
 P(\mathbf{x}; \vec k)
&\equiv&
 4 T(\mathbf{x}) \gamma_{\vec k}
 \simeq
 P^{(0)}(\vec k)
 +
 4
 \bigg(
 \mathbf{w}^T\mathbf{x}
 +
 \frac{1}{2} \mathbf{x}^T \mathbf{W} \mathbf{x}
 \bigg)
 \gamma_{\vec k}.
\end{eqnarray}
In the above formulae: \hfill\break
($i$) in a considered spin-orbital model (\ref{som}), the $P({\vec k})$
term is factorized into the coefficient $4T$ and a $\vec k$-dependent
term $\gamma_{\vec k}$ describing the dispersion; \hfill\break
($ii$) for the sake of clarity the following symbols were introduced:
$\mathbf{u}$ and $\mathbf{U}$ for the gradient and the Hessian of
$I(\mathbf{x})$, and $\mathbf{w}$ and $\mathbf{W}$ for the gradient
and the Hessian of $T(\mathbf{x})$, all at $\mathbf{x}=0$. \hfill\break
The above expansions were truncated at quadratic terms, so that the
corresponding variational function for a magnon energy has a quadratic
form,
\begin{equation}
 \omega_{\vec k}(\mathbf{x})
=
 I(\mathbf{x})
 +
 P(\mathbf{x}; \vec k)
 \simeq
 \omega_{\vec k}^{(0)}(\mathbf{x})
 +
 \Big( \mathbf{u}^T + 4\gamma_{\mathbf{k}}\mathbf{w}^T \Big)\mathbf{x}
 +
 \frac{1}{2} \mathbf{x}^T
 \Big( \mathbf{U}+4\gamma_{\mathbf{k}}\mathbf{W}\Big)\mathbf{x},
\end{equation}
that may be minimized to obtain
\begin{equation}
 \label{eq:app:delta_e_taylorUT}
 \omega_{\vec k}
 \simeq
 \omega_{\vec k}^{(0)}
 -
 \frac{1}{2}
 \Big( \mathbf{u}^T + 4 \gamma_{\mathbf{k}} \mathbf{w}^T \Big)
 \Big( \mathbf{U} + 4 \gamma_{\mathbf{k}} \mathbf{W} \Big)^{-1}
 \Big( \mathbf{u} + 4 \gamma_{\mathbf{k}} \mathbf{w} \Big).
\end{equation}
Note that as long as $I(\mathbf{x})$ and $T(\mathbf{x})$ may be
expressed analytically, so do $\mathbf{u}$, $\mathbf{U}$, $\mathbf{w}$,
$\mathbf{W}$, and, finally, $\omega_{\vec k}$ as well. Owing to this,
the above formula offers analytically the approximate results without
involving any further numerical minimization
(as opposed to the strategy used in the main part of the article).

If $J_1$ is perceived as a parameter in a generic form of a dispersion
relation $\omega_{\vec k} = 4J_1S(1-\gamma_k) + \ldots$, where all
other terms that may be introduced for better accounting of the
functional dependence such as $\propto J_3\cos(2k_a)$ are not written
explicitly, then its approximate value may be extracted directly from
Eq. \eref{eq:app:delta_e_taylorUT} as a coefficient in front of
$-4S\gamma_k$:
\begin{equation}
  \label{eq:J1}
 J_1^{\rm{app}}
 =
 J_{\lozenge}
 +
 \frac{1}{S}
 \bigg(
  \mathbf{u}^T
  \mathbf{U}^{-1}
  \mathbf{w}
  -
  \frac{1}{2}
  \mathbf{u}^T
  \mathbf{U}^{-1}
  \mathbf{W}
  \mathbf{U}^{-1}
  \mathbf{u}
 \bigg),
\end{equation}
where the second term captures the deviation form the frozen orbital
description.

In order to rationalize the reported value of $J_1$ for LaMnO$_3$ we
used Eq. (\ref{eq:J1}) together with the SVA parametrization.
The obtained value of the correction is equal to
$J_1^{\rm{app}}=-4.90\times10^{-3}J$, in fairly good agreement with the
value $-5.22\times10^{-3}J$ resulting from the fit. To avoid lengthy
formulae we do not present here a similar approach to determine $J_2$
and $J_3$, and restrict this analytic consideration to the SVA.

\section*{References}

\end{document}